# A Books Recommendation Approach Based on Online Bookstore Data

Xinyu Wei, Jiahui Chen, Jing Chen, Bernie Liu

## Abstract

In the era of information explosion, facing with complex information, it is difficult for users to choose information of interest, and businesses also need detailed information on ways to let the ad stand out. By this time, it is recommended that a good way.

We firstly by using random interviews, simulations, asking experts, summarizes methods outlined the main factors affecting the scores of books that users drew. In order to further illustrate the impact of these factors, we also by combining AHP consistency test, then fuzzy evaluation method, empowered each factor, influencing factors and the degree of influence come.

For the second question, predict user evaluation of the listed books from the predict annex. First, gave the books Annex labels, user data extraction score books and mathematical analysis of data obtained from spss user preferences and then use software to nearest neighbor analysis to result predicted value.

## 1. Introduction

Expansion in the information age, how to select interesting information and how to get information on all major outstanding issues is a significant problem. The recommendation is an important tool to resolve this contradiction, the products and applications of the Internet are widely used, such as in the areas related to the search topic recommendations, e-commerce has been widely used. We are given user basic behavior information from a famous online bookstore, such as read the history ,books ratings, user ID, books ID, book labels. According to the data, complete the following questions:

Analyze the effects on book remarks by users;

Design a model to predict the book remarks by users on their read books;

Recommend every user (attached in the file 'predict.txt') three books, which they have yet read before.

## 2.Analysis

### 2.1 Question one

For the evaluation factors influence the reader books are diverse, we can by asking experts, taking random interviews and simulation experiments, summarize and then draw book evaluation factors influence on readers. It could be summarized as summarized content, price and appearance these three main aspects. Then establishe two specific targets based on these three levels. In order to make the degree of influence factors more specifically reflected, can be used AHP factors given weight. Then, using fuzzy evaluation method especially fuzzy judgment matrix for the final objective evaluation. Constructe the indexes of an evaluation divided into excellent (100 points), good (75 points), medium (60 points), and poor (35 points) by the reader the survey scores. Then ghet the survey results, one by one the things to be quantified assessment from each evaluation index, derived composite score. And substituting into the authentication instance.

**2.2 Question two**

To predict user ungraded item, we use the evaluation scores for the user's preferences as a method of prediction. But there was a certain one-sidedness of the score by this method. However, it can be improve on this basis. Specific ideas are as follows: firstly, clear the user's preferences (can get the similarity from the labels on the read books and user scores of similar books collection by means of the angle cosine similarity), then find the target user preferences similarity to certain types of clusters that user's social circle. Target users' friends ratings of target prject and user preferences are combined through the use of ant colony algorithm, namely the use of a group of projects rated clustering prediction method first categories of users, and then calculate the predicted evaluation score books.

2.3 Question three

For the third question asked to recommend books, we can use collaborative filtering recommendation method: When the user books are recommended, we focused on his concern for the user to select the number and ratings as well as his own selection of books and scoring information, besides the use of similar measure to find the current user interest similar set of users. Then use the results from nearest neighbor mathematic to calculate the scores of neighbor users' hobby and interest. And then, predict the value of the current user who may be interested in the books, so a high degree of intelligence algorithms, he might be able to compare the new books of interest to users found to achieve a more intelligent and effective recommendation.

# 3.Assumption

(1) User preferences are established;

(2) the user's evaluation is true;

(3) each class of users are independent who will not be affected by other categories of users;

(4) User reviews are objective and true;

(5) the user's interest in reading seldom changes in a short time.

# 4.Symbols

| Symbol | Symbol Meaning |
|---|---|
| $C.R.$ | Consistency ratio |
| $B_i$ | Two indicators weight in an indicator |
| $C_i$ | Two indicators weight in the overall |
| $A_i$ | An index weight in over all |
| $R$ | Fuzzy relationship matrix |
| $R$ | Fuzzy relationship matrix elements |
| $R_{ij}$ | Users - Book scoring matrix elements |
| $M$ | The number of users |
| $N$ | The number of book projects |
| $u_j$ | The average similarity |
| $A$ | Similarity parameters |
| $V$ | Ants moving speed |
| $v_{amx}$ | Ants maximum speed |
| $Neigh_{i\times j}(r)$ | The square area from $r$ whose length is $s$ |
| $d(u_i,u_j)$ | Space distance between $u_i$ and $u_j$ |
| $sim(u_i,u_j)$ | Similarity of objects $u_i$,$u_j$ |
| $M$ | Number of properties of the object |
| $u_{i\times k}$ | The value of the k attribute of the user i |
| $p_p$ | Pick up probability |
| $p_d$ | Drop probability |
| $p_i$ | Class clusters |
| $c_i$ | Probability whether has been selected |
| $A$ | Threshold |
| $S$ | The total number of class clusters |
| $sim(u,c_i)$ | User u similarity of the class clusters $c_i$ |
| $D$ | Total project |
| $Center_{ci,r}$ | Users of cluster centers ratings to $r$ |
| $u_r$ | Users $u$ ratings for the project $r$ |
| $\delta_i$ | Density clusters of i |
| $N_i$ | The number of users within a class cluster $c_i$ |
| $T_u$ | Total number of users of e-commerce system |
| $U$ | Target users |
| $I_u$ | Project Space |
| $N_u$ | Ungraded items collection |
| $Y_u$ | Collection of items scored |
| $C$ | Nearest neighbor cluster collections |
| $V$ | Select number of the nearest neighbor cluster |
| $I$ | Target project |
| $U$ | Class clusters in the neighborhood of the project $i$ over the set of users rated |

# 5.Mathematic Models

**5.1 question one**

**5.1.1 Establish user rating impact indicators**

Through the investigation and the information from the reader's view , to figure out the impact factor score of books reader issue, we have an investigation on the relevant experts, librarians and readers.After inspecting evaluation of relevant online bookstore and reader feedback, through summary, we drew the following conclusions: The most important factor in terms of effect evaluation user book is the content of the book :

( 1 ) meets the needs of readers for content ;

( 2 ) For some non-academic books , the plot of the book is colorful or attractive ;

( 3 ) For a number of academic books , whether the contents of the book have a greater value in use , having the ability to meet academic needs of readers and the content is deep;

( 4 ) books are complete, with complete knowledge system architecture , content organization and structure of rational knowledge;

( 5 ) whether the content into appropriate sections or modules , in line with the internal logic of the system of books ;

( 6 ) the appearance rates of error conditions in the book.

Secondly, the larger one is the impact of the price of books .

(1 ) whether the price of books and books content and production values match the values;

( 2 ) the difference in price compared to similar books ;

( 3)whether is various other forms of the support compared with similar content , the price is appropriate ;

( 4 ) whether the production of materials and price of books in line ;

Then evaluate the factors that affect the appearance of the book :

( 1) Book printing conditions ( text and graphics are clear ) ;

( 2 ) whether the content rich ;

( 3 ) whether the unique book cover is appropriate , pleasing ;

( 4 ) whether books are fine .

The index made into the form:

| First index | Second index | Evaluation Criteria |
|---|---|---|
| content | Meet the needs of the reader | Meet the needs of readers content |
| | Abundance | The plot of the book is rich |
| | Integrity | Books contents are complete |
| | Definition | Reasonable structure, associated with a clear knowledge |
| | Academic contribution | Whether they have a greater content of value |
| | Correct rate | Error condition occurs on book |
| | Readers interest | Whether readers are interested in the content of books |
| Price | Value for money | Whether the price of the book valued and consistent |
| | Compared to similar books | Are similar books compared to the price difference; |
| | Compared with other carrier | Different forms of carrier comparison with similar content |
| | Production of raw materials | Production of materials and price matching books |
| Outlook | Print | Books printed case |
| | Carrier Forms | Carriers are diverse forms of content |
| | Written performance | There are pleasing performance in book form |
| | Production process | Books are exquisite craftsmanship |

In addition, there are factors such as collectible value, meaning, the historical value of books, etc., has little effect on this, so they can be ignored. For content, price, appearance, etc. level indicators and numerous secondary indicators, to demonstrate each indicator the degree of impact on the user rating specifically, we used the AHP and fuzzy comprehensive evaluation method to determine the weight of each evaluation.

### 5.1.2 Analysis Hierarchy Process

Before the establishment of fuzzy evaluation model, we need to use every level of AHP to give comparative evaluation matrix, and then were calculated for each level of the heavy weight of each index and its one-off test. Before you can use the weights derived fuzzy evaluation model. Divide into the level

Through data access and visited by readers and ask experts and other methods, we arrive at the following levels of indicators:

| First index | Second index |
|---|---|
| Content | Meet the needs of the reader, richness, integrity, clarity, academic contributions, the correct rate, the reader interested |
| Price | Cost-effective, compared with similar book ratio, compared with other carriers, production of raw materials |
| Outlook | Printing, the carrier, writing performance, production process |

(2) the establishment of first index of judgment matrix and calculate the weight

Establish an index contrast matrix, and through the use of Matlab-time inspection to determine whether that CR is less than 0.10, if it is, then continue the transformation of matrix contrast, until CR is less than 0.10. If so, we determine weights Q. After testing and operations, comparison matrix and the weight are as follows: (op-program see Appendix I)

| | content | price | outlook |
|---|---|---|---|
| content | 1 | 2 | 6 |
| price | 1/2 | 1 | 3 |
| outlook | 1/6 | 1/3 | 1 |
| Weights | 0.6 | 0.3 | 0.1 |

For consistency of judgment matrix, each column after normalizing is one of the corresponding weights. For non-consistency of judgment matrix, each column after normalization is similar to its corresponding weight, in this strike n column vector arithmetic mean as the final weights. Specific formula is:

$$w_i = \frac{1}{n}\sum_{j=1}^{n}\frac{a_{ij}}{\sum_{k=1}^{n} a_{kl}}$$

So the judgment matrix can be drawn as follows:

$$A = \begin{bmatrix} 1 & 2 & 6 \\ 1/2 & 1 & 3 \\ 1/6 & 1/3 & 1 \end{bmatrix}$$

(1) consistency check (source code, see Appendix III):

The first step: Calculate the consistency index

$$C.I. = \frac{\lambda_{\max} - n}{n-1}$$

Step two: look-up table to determine the corresponding average random consistency index RI

Richard corresponds to a value of 0.52 R.I.

The third step: the consistency ratio C.R. is calculated and judged, the formula is:

$$C.R. = \frac{C.I}{R.I}$$

When C.R. <0.1, the consistency of judgment matrix is considered acceptable, CR> 0.1, consider the consistency of judgment matrix does not meet the requirements, the need to re-amend the judgment matrix. Upon examination, we found that the value of CR is in the 0.1, it can be considered to determine the matrix is in line with the weight distribution. Similarly: According to the above method, we can draw two indicators account for level indicators where the weights Bi, where the level is then multiplied by a matrix and the overall weight of Ai, each of the two indicators can be drawn representing the overall weight of Ci, That

$$C_i = A_i * B_i$$

Finally, each of the two indicators can be drawn representing the overall weight is:

| Firsr index | Second index | weight |
|---|---|---|
| content(0.6) | Meet the needs of the reader | 0.2 |
| | Abundance | 0.1 |
| | Integrity | 0.05 |
| | Definition | 0.05 |
| | Academic contribution | 0.05 |
| | Correct rate | 0.05 |
| | Readers interest | 0.1 |
| price(0.3) | Value for money | 0.1 |
| | Compared to similar books | 0.1 |
| | Compared with other carrier | 0.05 |
| | Production of raw materials | 0.05 |
| outlook（0.1） | Print | 0.03 |
| | Carrier Forms | 0.02 |
| | Written performance | 0.04 |
| | Production process | 0.01 |

**5.1.3 Fuzzy Comprehensive Evaluation Method**

The right to use AHP to determine weight is a lot of subjectivity, it requires the use of fuzzy judgment matrix for the final objective evaluation of survey rated by readers that 1 point, 2 points, 3 points, 4 points, 5 points. Then we got survey results, one by one to be quantified assessment of things from each evaluation index, which is determined from the single factor was rated things look fuzzy subset of grade of membership ($R|u_i$), then get fuzzy relationship matrix:

$$R=\begin{bmatrix} R| & u_1 \\ R| & u_2 \\ \cdots & \\ R| & u_p \end{bmatrix}=\begin{bmatrix} r_{11} & r_{12} & \cdots & r_{1m} \\ r_{21} & r_{22} & \cdots & r_{2m} \\ \cdots & \cdots & \cdots & \cdots \\ r_{p1} & r_{p2} & \cdots & r_{pm} \end{bmatrix}_{p.m}$$

By element validation and comparison of fuzzy relationship matrix findings and validated into instances found such a method of assigning weights derived from actual books online scores and evaluation scores are basically the same, so it can be concluded papers weight distribution is reasonable. So the first question can be summarized as: the main factors influencing book readers evaluation scores are the content, price, appearance, content of which impact on the reader's score, followed by the price, and finally the appearance. In the specific secondary indicators, the biggest factor is the degree to meet the needs of readers, followed by readers' interest, richness and so on. Finally, a list sorted is listed in the form according to the impact from high to low, as follows:

| Influencing index | Influencing degree |
|---|---|
| Meet the needs of the reader | 0.2 |
| Abundance | 0.1 |
| Integrity | 0.1 |
| Definition | 0.1 |
| Academic contribution | 0.1 |
| Correct rate | 0.05 |
| Readers interest | 0.05 |
| Value for money | 0.05 |
| Compared to similar books | 0.05 |
| Compared with other carrier | 0.05 |
| Production of raw materials | 0.05 |
| Print | 0.04 |
| Carrier Forms | 0.03 |
| Written performance | 0.02 |
| Production process | 0.01 |

## 5.2 The second question

### 5.2.1 Method one, calculated using the user's preferences to predict scores [3]

Rationale: It is known to us that user rating for the same type of books are basically the same, and the same judgment on whether the book is the first of a class label are the same. First with the table in the user data in the user's score will be rated books where the first label to find out. Find another user score data to the user, for each user to use the book to score the first label to find out that it has scores of similar books, then record books ID, the total number of tags and scores. Reuse Nearest Neighbor Analysis Spss be drawn predictive value.

User ID is 7245481, 794171 book ID is an example:

Because the data is large, we listed only part of the data.

Third table

| Similar books | Score |
|---|---|
| 199155 | 2 |
| 232671 | 4 |
| 318311 | 4 |
| 479614 | 4 |
| 997265 | 4 |

Finally, we used Spss software to analyzed the nearest neighbors, drew the following diagram:

案例处理汇总

| | | N | 百分比 |
|---|---|---|---|
| 样本 | 训练 | 5 | 100.0% |
| | 保持 | 0 | .0% |
| 有效 | | 5 | 100.0% |
| 已排除 | | 0 | |
| 总计 | | 5 | |

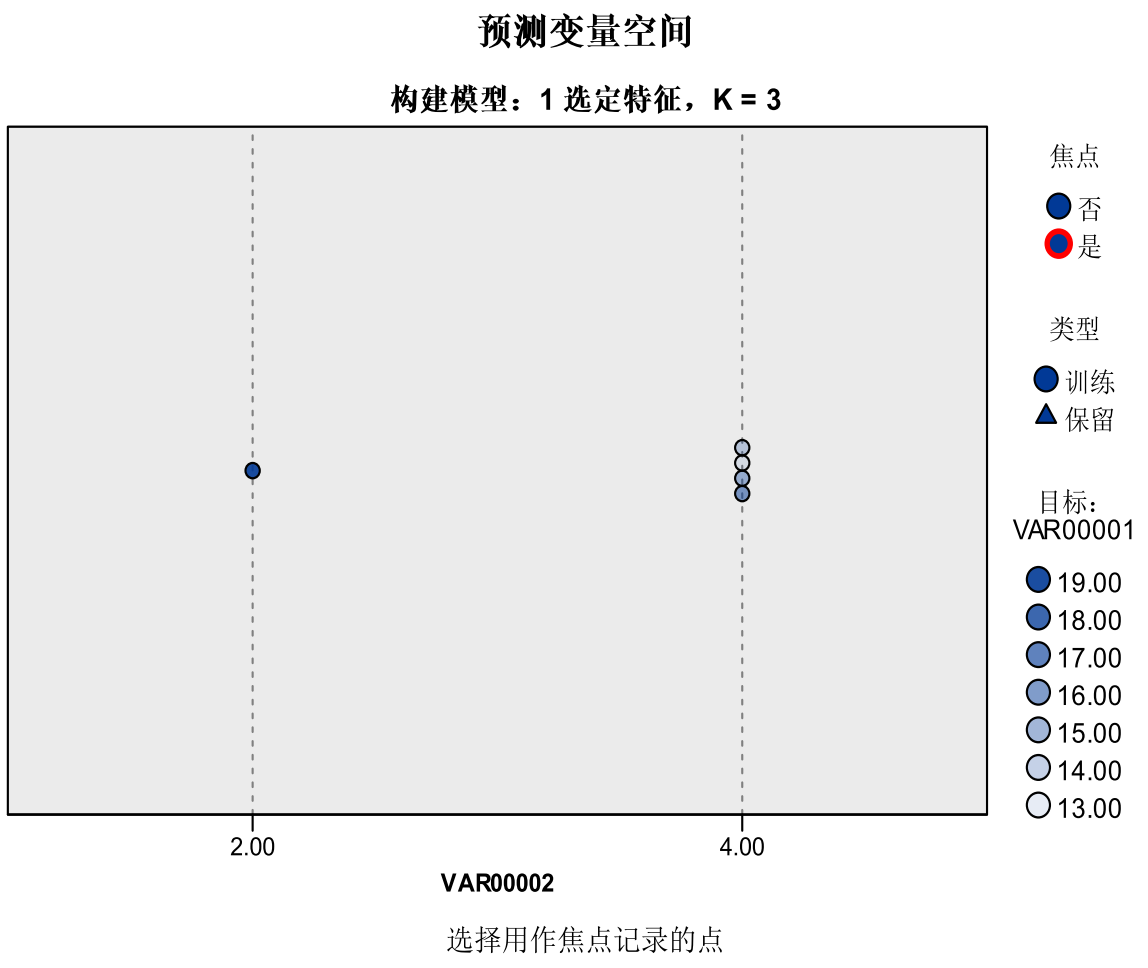

Therefore, predictive value of 4 points can be drawn. Other cases and so on, basically the same. We got the conclusion that each user's preferences on the need to predict the books, just from the preferences, the predictive value of 4 points. But only from the preference is too one-sided, and also taking into account the user social circle, applying the use of the ant colony clustering project score prediction.

### 5.2.2 Method Two: ant clustering Item Rating Prediction

The first step: the basic principles of ant colony clustering [4]: prepare clustering objects randomly distributed in two-dimensional coordinates of the object, the similarity with ants is measured by the probability of converting this function to convert the similarity picked up, moved, or probability is down, after several iterations, similar data can be clustered together.

(1) set of data objects $u_i$ is found within its domain objects $u_j$ which represents the average similarity

$$\left(f_{u_i}\right)=\max\left\{0,\frac{1}{s^2}\times\sum_{u_j\in Neigh_{i\times j}(r)}[1-\frac{d(u_i,u_j)}{a(1+(v-1)/v_{\max}}]\right\}$$

Inside, $\alpha$ is similarity parameter, $v$ is ant speed, $v_{\max}$ maximum speed for the ants, $Neigh_{i\times j}(r)$ to impose a location area $r$ of a square side whose length is $s$, and $d(u_i,u_j)$ is the space for the data object $u_i$ and $u_j$ distance. In this paper, we used the cosine law metric: [5]

$$d(u_i,u_j)=1-sim(u_i,u_j) \quad (2)$$

$sim(u_i,u_j)$ is the similarity of data objects $u_i$ and $u_j$, the similarity with the cosine angle between two vectors to measure the object attributes, then

$$sim(u_i, u_j) = \frac{\sum_{k=1}^{m} (u_{i \times k} \bullet u_{j \times k})}{\sqrt{\sum_{k=1}^{m} (u_{i \times k})^2 \sum_{k=1}^{m} (u_{j \times k})^2}} \quad (3)$$

Is the number of properties of an object, for the first attribute value of the object, the more similar the data object, the more the value of close to 1, otherwise close to 0.

(4)the pick up and drop probability

$$p_p = \left(\frac{k_1}{k_1 + f(u_i)}\right)^2 \quad (4) \qquad p_d = \begin{cases} 2f(u_j) \, if & f(u_j) < k_2 \\ 1 \quad if & f(u_j) \geq k_2 \end{cases} \quad (5)$$

In formula, $p_p$、$p_d$ were picked up and discarding probability, as $k_1, k_2$ are the threshold constant, if $f(u_j)$ is bigger, $p_d$ is smaller, $p_p$ is bigger, and vice versa.

Step two: user clustering generation [3]: Currently, there are many clustering algorithm for research and application. Among them, the methods applied mostly are $k-$ clustering, neural network clustering, fuzzy clustering. These drawbacks of clustering algorithms are the order of over-reliance on input elements which needs manual specify the number of clusters and human center and irreversible process of clustering , so it is not suitable for complex group. The advantages of ant colony clustering is the clustering process visualization, the number of clusters in the clustering process can be automatically generated and can achieve a fully distributed control. In terms of its self-organization, to expansion, and robustness are better than traditional clustering algorithms. Steps are as follows:

(1) to all users in any two-dimensional spatial coordinates hash;

(2) using Equation (1) to (5) to configure relational databases, achieve users clustering and have multiple iterations;

(3) gather the similar user together as a class, the output of the clustering results and $D = \{d_1, d_2, d_3, \ldots d_i, \ldots d_n\}$, $n$ is the number of clusters.

As shown: for details in Appendix

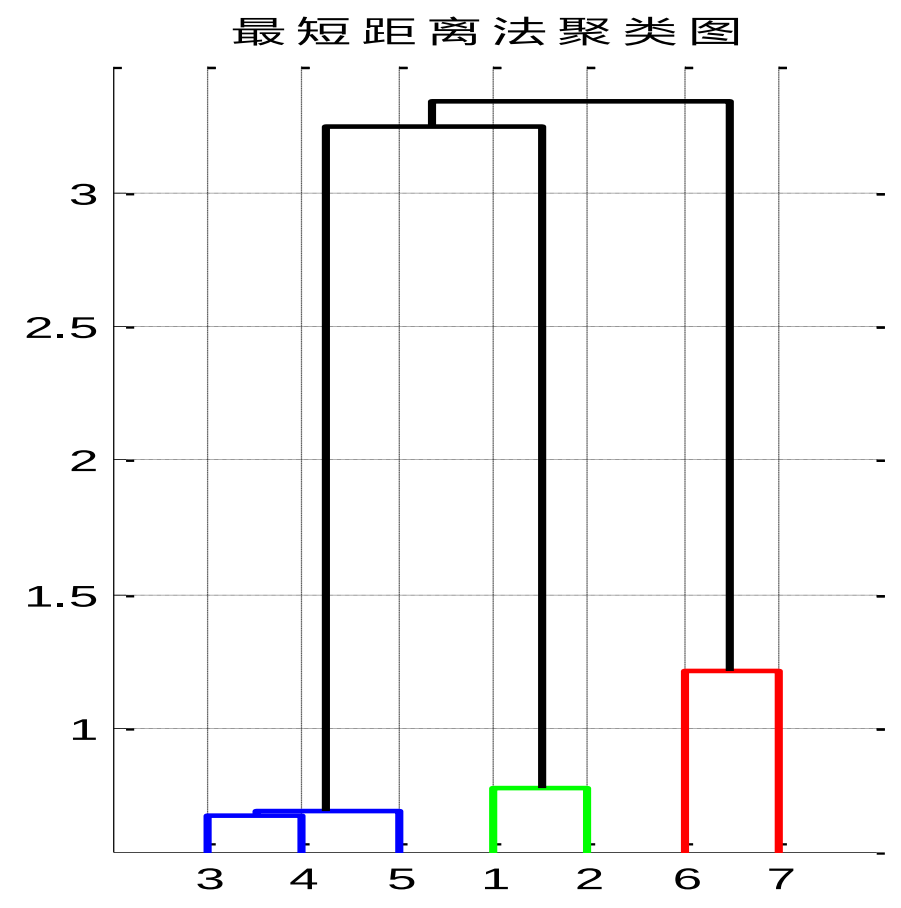

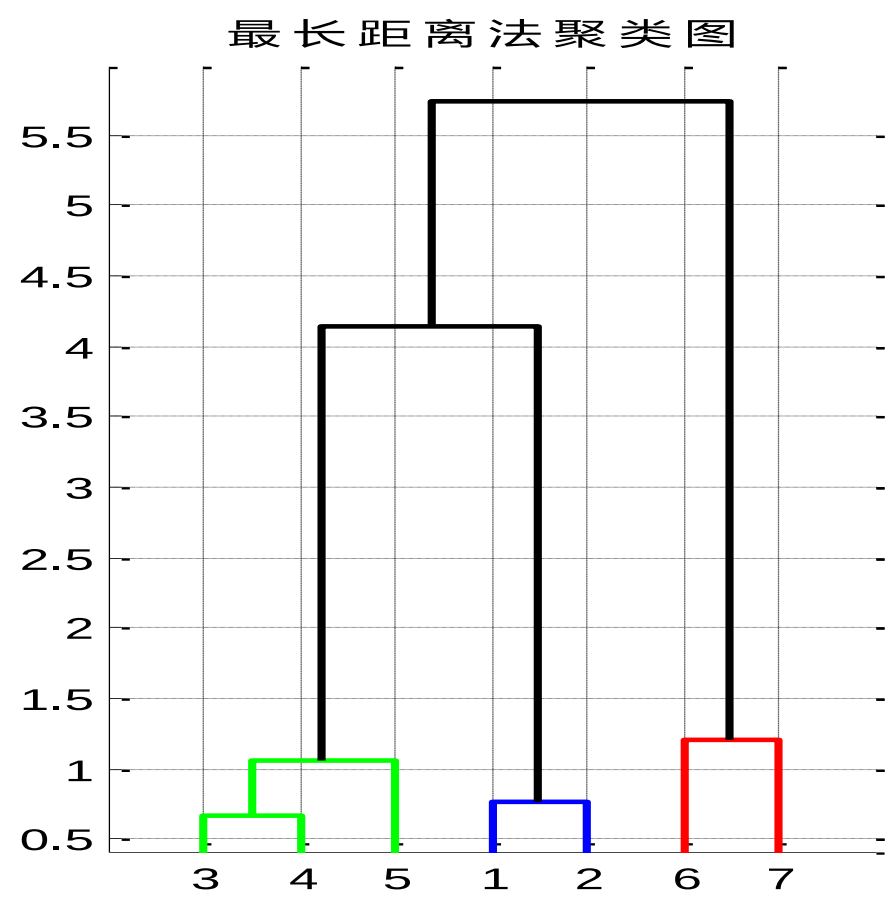

Rating prediction:

(1) neighbors class cluster selection [6]

$$p_i = \frac{\delta_i \bullet sim(u,c_i)}{\sum_{i=1}^{s} \delta_i \bullet sim(u,c_i)} \quad (6)$$

Wherein $p_i$ is the class clusters , $c_i$ is the probability whether is selected, if it is larger than the threshold $\lambda$ it will be selected, otherwise abandoned, $s$ is the total number of cluster class, $sim(u,c_i)$ is the similarity of the target user $u$ and class cluster $c_i$ , measured by the user and the cluster center distance, i.e.:

$$sim(u,c_i) = \sqrt{\sum_{r=1}^{d} \left| center_{c_i,r} - u_r \right|^2} \quad (7)$$

$d$ represents the total number of projects $center_{c_i,r}$ is the cluster center users ratings for project $r$ , $u_r$ is users $u$ ratings for project $r$ , $\delta_i$ is the class cluster density, expressed as:

$$\delta_i = \frac{N_i}{T_u} \quad (8)$$

Among them, $N_i$ is the number of users from class cluster $c_i$ , $T_u$ is the total number of users of all classes within the e-commerce systems.

Provided that target user is $u$ ,project space $I_u$ for the collection of items which is not rated, has a collection of items for the score $N_u$ .rated project collaboration is $Y_u$ ,

Then $N_u = I_u - Y_u$, provided that the nearest neighboring class cluster collaboration of selected users is $C = \{c_1, ...c_i, ...c_v\}$ , $v$ is the number of the selection of nearest neighbors class clusters, and as shown, detailed see Appendix II

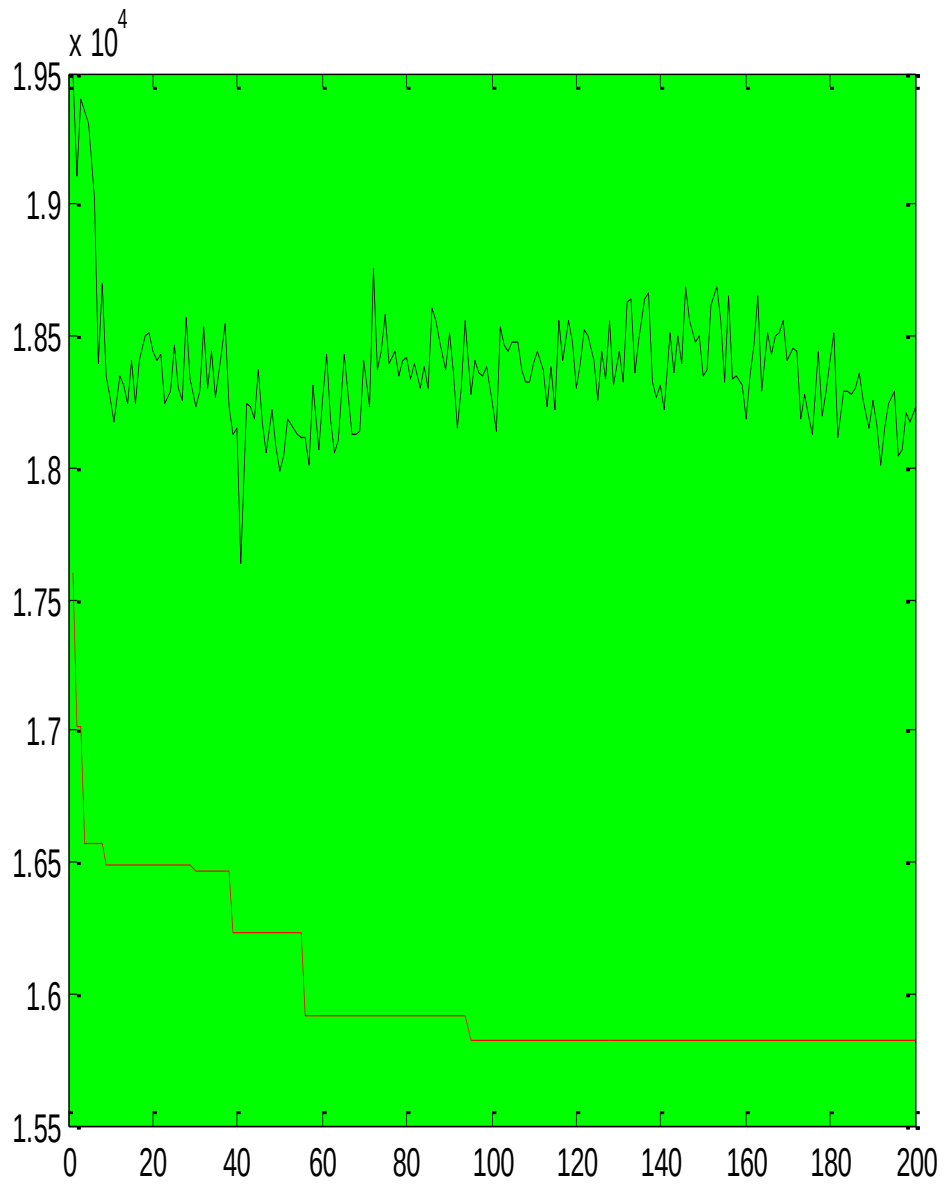

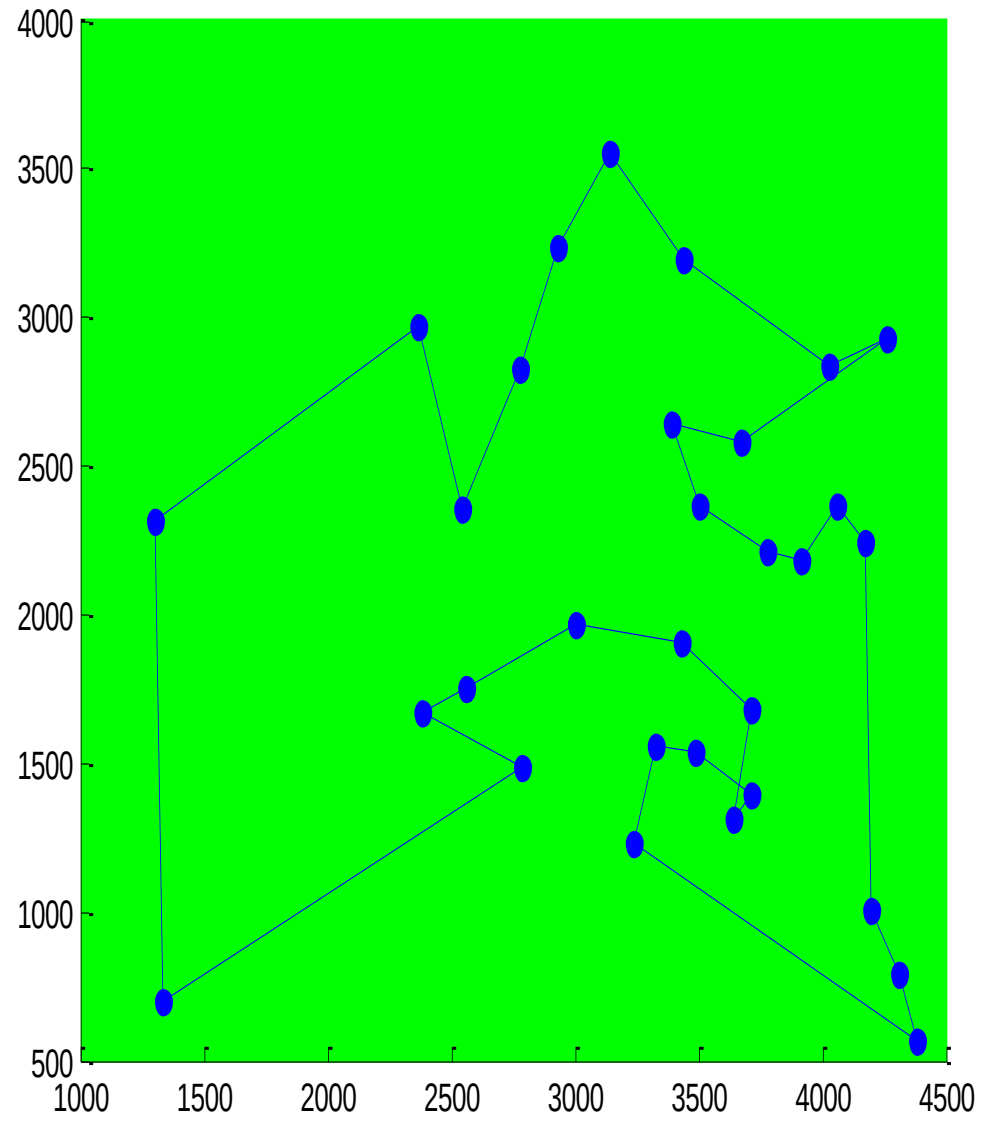

(2) scoring method [5]

For target projects $i, i \in N_n$, the collection of users who from the neighboring class cluster gave rates to project $i$, is shown as $U = \{u_i, u_2, \ldots u_s \ldots u_n\}$ ,and the method by which assess user $u$ gave scores to the goal of the project $i$ as follows:

$$p_{u,i} = \begin{cases} \sum_{u_\partial \in U} sim(u, u_\partial) \times r_{u_\partial, i} \; if \quad U \neq \varnothing \\ 0 \; if \quad U = \varnothing \end{cases}$$

Among them, $sim(u, u_\partial)$ is the degree of acquaintance with the user $u$ and users $u_\partial$ from class cluster $r_{u_\partial, i}$ is user $u_\partial$ assessment to target project $i$

(3) Conversion rates

Following the above process, there is a score value for most projects in user-project rating database. That is user $u$ ratings to project $i$ can be converted as follows:

$$R_{u,i} = \begin{cases} r_{u,i} \; if \quad user \, u \quad rate \quad item i \\ p_{u,i} \; if \quad user \quad u \quad not \, rate \quad item i \end{cases}$$

Finally, the predicted value was obtained from the following circles:

| userID | bookID | | | | | |
|---|---|---|---|---|---|---|
| 7245481 | 794171 | 381060 | 776002 | 980705 | 354292 | 739735 |
| score | 5 | 3 | 2 | 4 | 3 | 1 |
| userID | bookID | | | | | |
| 7625225 | 473690 | 929118 | 424691 | 916469 | 235338 | 793936 |
| score | 4 | 3 | 1 | 3 | 2 | 5 |

| userID | bookID | | | | | |
|---|---|---|---|---|---|---|
| 4156658 | 175031 | 422711 | 585783 | 412990 | 134003 | 443948 |
| score | 2 | 4 | 5 | 3 | 1 | 3 |
| userID | bookID | | | | | |
| 5997834 | 346935 | 144718 | 827305 | 219560 | 242057 | 803508 |
| score | 4 | 3 | 2 | 5 | 1 | 3 |
| userID | bookID | | | | | |
| 9214078 | 310411 | 727635 | 724917 | 325721 | 105962 | 234338 |
| score | 1 | 4 | 5 | 2 | 2 | 3 |
| userID | bookID | | | | | |
| 2515537 | 900197 | 680158 | 770309 | 424691 | 573732 | 210973 |
| score | 5 | 3 | 4 | 2 | 1 | 3 |

5.2.3 The weighted average method

Use the weighted average, arrive at a final prediction score (rounded) expressed as follows:

$$y = \frac{y_i + y_j}{2}, y \in N^*$$

The end result is as follows:

| userID | bookID | | | | | |
|---|---|---|---|---|---|---|
| 7245481 | 794171 | 381060 | 776002 | 980705 | 354292 | 739735 |
| score | 5 | 4 | 3 | 4 | 4 | 3 |
| userID | bookID | | | | | |
| 7625225 | 473690 | 929118 | 424691 | 916469 | 235338 | 793936 |
| score | 4 | 4 | 3 | 4 | 3 | 5 |
| userID | bookID | | | | | |
| 4156658 | 175031 | 422711 | 585783 | 412990 | 134003 | 443948 |
| score | 3 | 4 | 5 | 4 | 3 | 4 |
| userID | bookID | | | | | |
| 5997834 | 346935 | 144718 | 827305 | 219560 | 242057 | 803508 |
| score | 4 | 4 | 3 | 5 | 3 | 4 |
| userID | bookID | | | | | |
| 9214078 | 310411 | 727635 | 724917 | 325721 | 105962 | 234338 |
| score | 3 | 4 | 5 | 3 | 3 | 4 |
| userID | bookID | | | | | |
| 2515537 | 900197 | 680158 | 770309 | 424691 | 573732 | 210973 |
| score | 5 | 4 | 4 | 3 | 3 | 4 |

5.3 Third Question

For the third question about books recommendation, we used collaborative filtering recommendation method: When the user recommend book based on his concerns to select the number of users and ratings of books and his own choices and score information, the use of similar measure to find the current user interest similar set of users. Then use the interest weighted nearest neighbor loving neighbor users to predict the value of the current user may be interested in the books, so a high degree of intelligence algorithms might be able to compare the new books of interest to users found to achieve a more intelligent and effective recommendation.

5.3.1 recommended steps

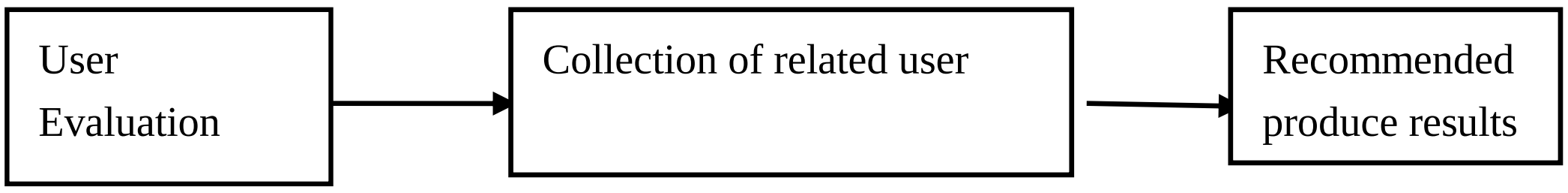

Beginning of the main steps and ideas are made:

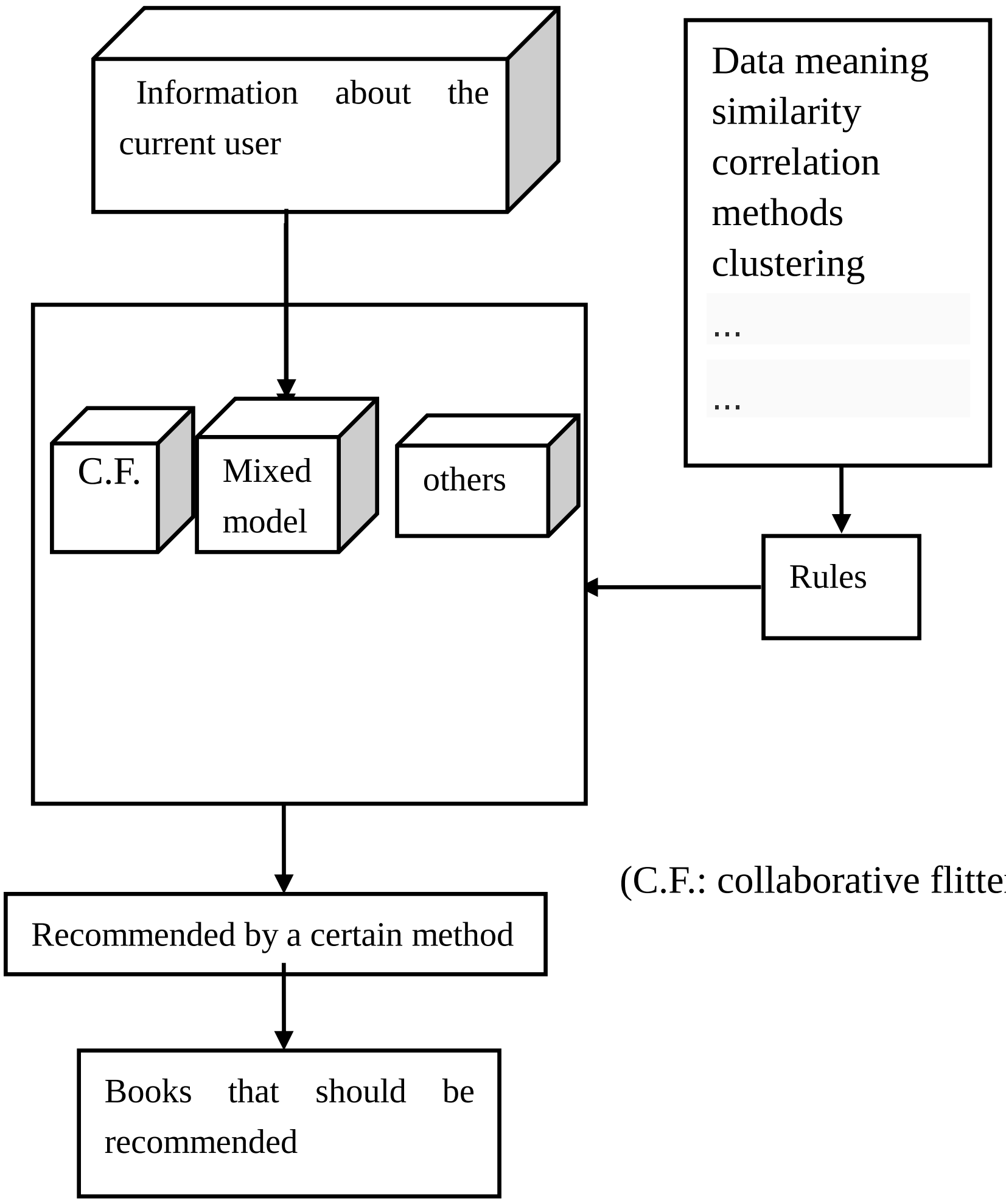

(C.F.: collaborative flitters recommendation)

Important summary, collaborative filtering method can be summarized as [6]:
(1) The user is classified according to their interests;
(2) the different information user evaluates includes the user's interest information;
(3) User evaluation of information to an unknown will be similar to similar (interest) User rating. This three regulations forms the basis of the collaborative filtering system.

### 5.3.2 score indication

The traditional collaborative filtering algorithm input data is the user-item ratings matrix of m × n, which M represents the number of users, and n represents the number of items.

| *User* | *Item* 1 | *Item* 2 | *Item* 3 | *Item* 4 | *Item* 5 | *Item* n |
|---|---|---|---|---|---|---|
| *User*1 | $R_{11}$ | $R_{21}$ | $R_{31}$ | $R_{41}$ | ... | $R_{n1}$ |
| *User*2 | $R_{12}$ | $R_{22}$ | $R_{32}$ | $R_{42}$ | ... | $R_{n2}$ |
| *User*3 | $R_{13}$ | $R_{23}$ | $R_{33}$ | $R_{43}$ | ... | $R_{n3}$ |
| *User*4 | $R_{14}$ | $R_{24}$ | $R_{34}$ | $R_{44}$ | ... | $R_{n4}$ |
| ... | ... | ... | ... | ... | ... | ... |
| *User* m | $R_{1\mathrm{m}}$ | $R_{2m}$ | $R_{3m}$ | $R_{4m}$ | ... | $R_{nm}$ |

Combining this passage can be concluded that the $R_{ij}$ can be 0,1,2,3,4,5 represent the degree of user preferences for this book (from 0 to 5 points, increase by degree of be fond of), while classification Numbers can also be used to indicate the user's preferences in the project. For those books which are not rated , can be replaced with a zero.

According to user - book score matrix, each user's score or score of every book, can be described through the corresponding row/column. Making the process of counting and analyzing easier.

### 5.3.3 score differences between user information entropy based similarity measure methods [7]

The principle of collaborative filtering algorithm recommendation is finding out congenial neighbor sharing the similar interest with current user. And in the topic, also includes the user concern of other users, through near neighboring user and the current user pays close attention to the evaluation of the current user recommend books to purchase customers.

In neighboring users (that is, with similar books hobby user) the choice of method is: calculating the similarity of current user and all other users in the recommendation system, according to the calculated similarity of the results, from big to small, in turn, sorting, choose the first K users as neighbor users collection, which will be intersected with the collection that the current user attention constituted, and the collection that we get is related user collection. The choice of similarity measure method for recommendation precision has a crucial influence, in order to improve the persuasion, we adopted similar measure based on information entropy method which is different from ant colony to work out question two, by comparing the results of two methods, to increase the accuracy of the results and persuasive.

We chose the similarity measure based on the score difference between user information entropy method of similarity measure (NWDE) for the following reasons: considering the user attention circle between the size of the intersection, a kind of weighted information entropy is used to measure the similarity of users score, score disparity between users by calculation, at the same time the method without the users or books or other attribute information, on the basis of the given topic data sparseness degree, relieve the traditional similarity measure methods which will increase or decrease too much similarity between users, improve the accuracy of the recommendation. Compared with some traditional collaborative filtering algorithm, the most used for Pearson correlation, Spearman correlation cosine similarity, while they consider the readers of the diversity of evaluation standard, however, these similarity measure methods in the collaborative filtering system also has the following disadvantages: (1) in high-dimensional sparse data, the size of the circle intersection that users pay close attention to is much smaller and inconsistent, the traditional similarity measure methods cannot solve such situation, the results are more likely to decrease or increase the real similarities between users. (2) affected by the factors such as data sparse, books recommended precision is low, which remains to be further improved. [7]

For a given data information, the information entropy calculation formula is:

$$H(x) = \sum_{i=1}^{n} p(a_j) \log_2 \frac{1}{p(a_i)}$$

Among them, n stands for books set X tag number, p (ai) represents the probability of the i class labels in X. Information entropy indicates that the greater the book tag is more discrete, the smaller the information entropy indicates that tag more gathering. When X probability of labels in the same age, the maximum information entropy, When there are only one X classification, take the minimum information entropy.

First: compare the score differences between two users. Assumes the shared collection of user i and j common rating is I, ratings data of user i and j are respectively $U_{\mathrm{i}} = \{R_{Ui,I1}, R_{Ui,I2}, R_{Ui,I3}, ..., R_{Ui,In}\}$ and $U_{\mathrm{i}} = \{R_{Ui,I1}, R_{Ui,I2}, R_{Ui,I3}, ..., R_{Ui,In}\}$

Then the two users rated degree of difference data $Diff(U_i, U_j)$ is:

$$\begin{aligned} Diff(U_i, U_j) &= \{R_{Ui,I1} - R_{Uj,I1}, R_{Ui,I2} - R_{Uj,I2}, ..., R_{Ui-In} - R_{Uj,In}\} \\ &= \{d_1, d_2, d_3, ..., d_n\} \end{aligned}$$

Then: weighted information entropy.

Use the formula of information entropy to calculate user rating degree of difference $Diff(U_i, U_j)$ 's Entropy

$$H\left(Diff\left(U_i,U_j\right)\right)=\sum_{i=1}^{n}p(d_i)\log_2\left(\frac{1}{p(d_i)}\right)$$

$$=-\sum_{i=1}^{n}p(d_i)\log_2 p\left(d_i\right)$$

Here, $P(d_i)$ represents the probability $d_i$ that in the $Diff(U_i,U_j)$. When the difference in the degree of entropy User rating is 0, then it is clear that both users ratings basically exactly the same, that similarity is the highest; Conversely, the higher the entropy of information, indicates that no two users rated more similar degrees.

In calculating the entropy $H(Diff)$ of score differences $Diff(U_i,U_j)$, we take the size reflected the user similarity degree, which is various, of the ratings of the difference di into account. The greater the di is, the reflecting differences in the user should be bigger. Thus, the weight are made the adjustments to the formula, when a weight is applied to information entropy | di |; in addition, considering the size of the intersection of two other users attention to the rest, we added a weight 1 / n as a symbol in the formula to increase the size of the intersection between the two weights (n is the size of the set intersection of these two users concerned users,while n is larger, the similarity of the user is higher, the difference corresponding to the rating information entropy is smaller). At this point, we can draw the weighted entropy difference (WDE (Ui, Uj)) of user i and j, which is calculated as:

$$WDE(U_i,U_j)=-\frac{1}{n}\sum_{i=1}^{n}p(d_i)\log_2 p(d_i)?|\mathrm{d}_i|$$

For current users a, work out the WDE of other users and a, before getting a vector WDE which can reflect the rating difference degree to books form user a and others.

The third step: $WDE_{Uai}$ is normalized to [0,1] matrix. By the equation, $WDE_{Ua}$ is in the range of zero to infinity, and therefore need to be normalized. Meanwhile, the greater the WDE（$U_i$,$U_j$）represents the greater the user differences. Therefore, we use the following electrode the value of the linear model to normalize elements from $WDE_{Ua}$: (see Annex)

$$NWDE_{Ua}[i]=\frac{Max(WDE_{Ua})-WDE_{Ua}[i]}{Max(WDE_{Ua})-Min(WDE_{Ua})}$$

Wherein NWDE [i] greater, indicates that the higher the similarity between users. The resulting scatter plots and line chart is below:

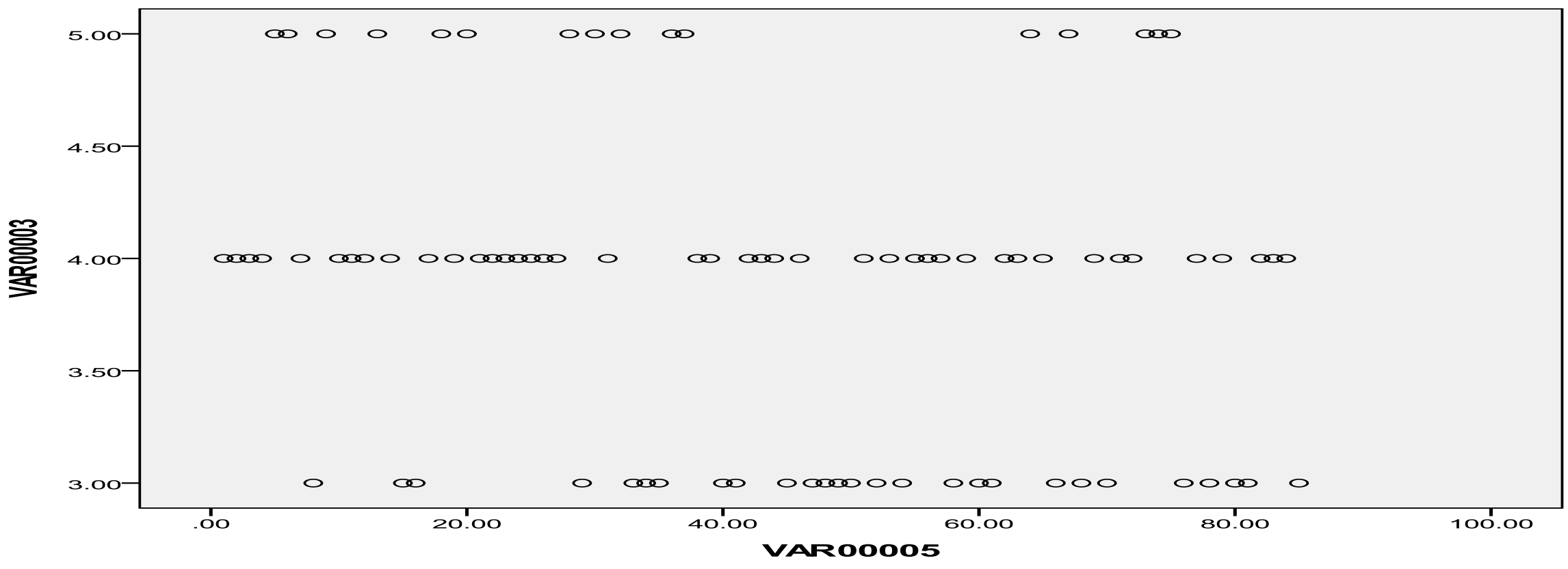

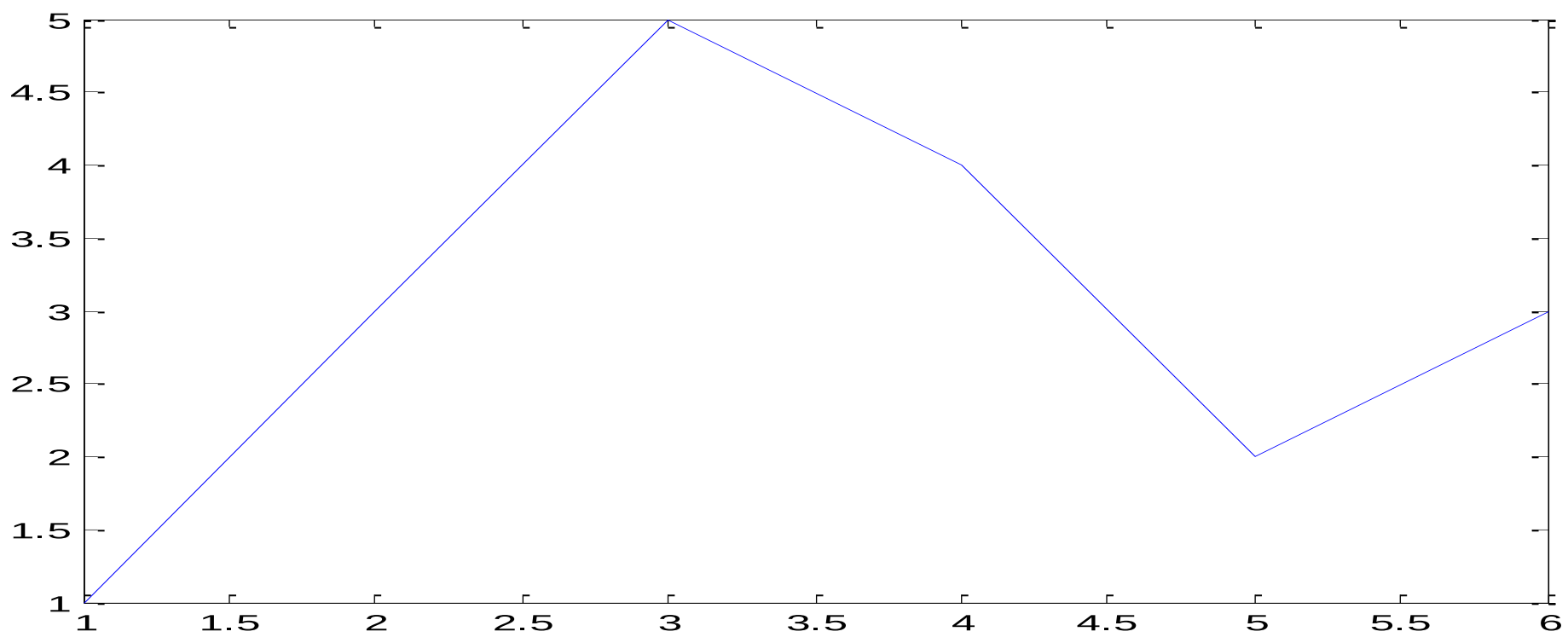

Use the given data from the title, the method of this paper and a computer simulation method, the similarity of the results of different methods of distribution in MATLAB software chart is drew as follows (which points represent different colors of different approximation calculation method ) [8]:

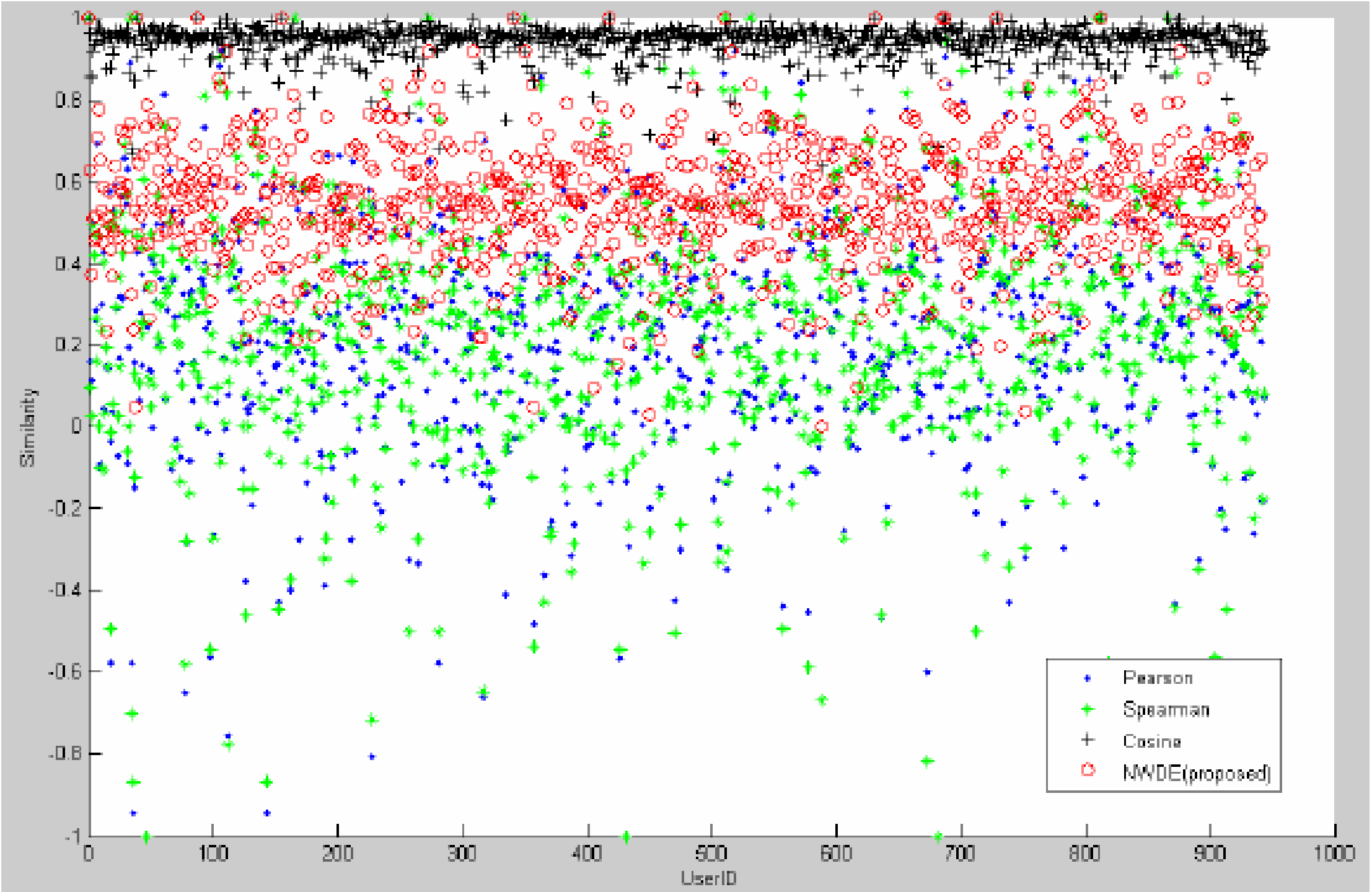

After calculating the degree of similarity among current users and other users, the use of prediction methods previously mentioned article is carried on the books of user score prediction:

$$P_{a,i} = \overline{R_a} + \frac{\sum_{u \in KNB} Sim(a,u)?(R_{u,j} - \overline{R_u})}{\sum_{u \in KNB} (Sim(a,u))}$$

Wherein $p_{a,i}$ represents the user a of a predictive score for book i, KNB indicates similar user collection of user a, $\overline{R_a}$ and $\overline{R_u}$ on behalf of the user u and the average of all scores in the score on all books, $R_{u,j}$ means that the user u giving score on Book i, Sim (a, u) represents the degree of similarity between two users interested in reading, in such algorithm , the value of Sim (a, u) is the value after normalization of $NWDE_{Ua.}$ Finally, the sort of value $R_{u,j}$ take forecast for the current user, the highest score in the first three books is the recommended to book.

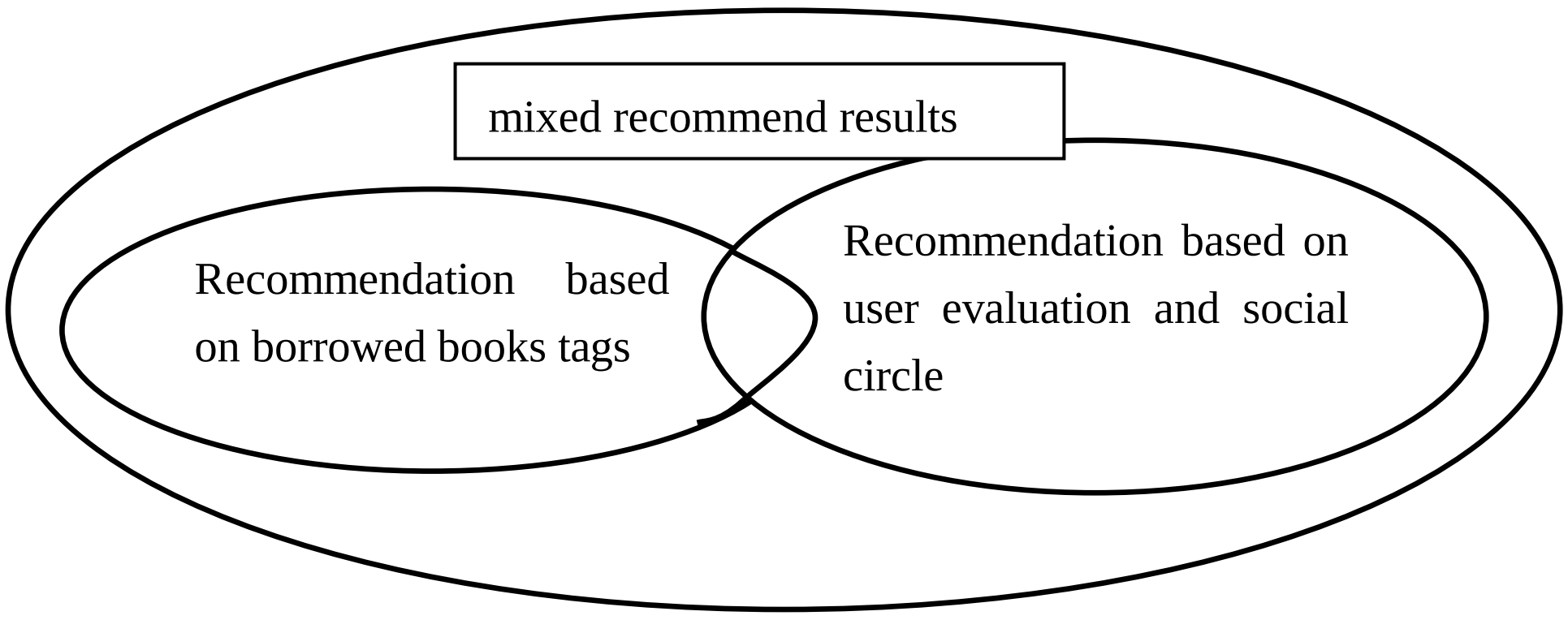

### 5.3.4 The following are the books recommended model [3]

Variability of user interest. user preferences are not always on the same book, with the accumulation of knowledge and change the living environment in which to read the books category will change. Such as college freshmen, first entered the school, the time is adequate, would like to see in some domestic pearl, books magazines and the like, but the junior and senior students under as fieldships, graduate job hooking and other pressures may be more like to see some employment and entrepreneurial guidance. In this case, to accurately recommendation is difficult. [9]

Available time by the number of books borrowed book can reflect this span, to a certain extent, a certain book recommended limits.

For recommended results through user recorded in question two: First calculate similar collections, sort of like a set of results for each recommendation element, according to the number of books to borrow from small to large.

For the recommend results of book tag in question three: the same number of books to borrow in accordance sort the results after each neighbor, the result is stored in a orderly results, sorting methods above. Because of the above two types of scores of different types can be determined by the results of the relationship between a factor of two, so that they are in the same order of magnitude, the final result is stored sequentially repeated to take into the final recommendation results.

Recommended mixing is used herein as the test result obtained before the two were the first row after the first method good order, and then interleaved by a second alternate method to arrive at the final recommendation result.

Recommended brief flowchart:

:

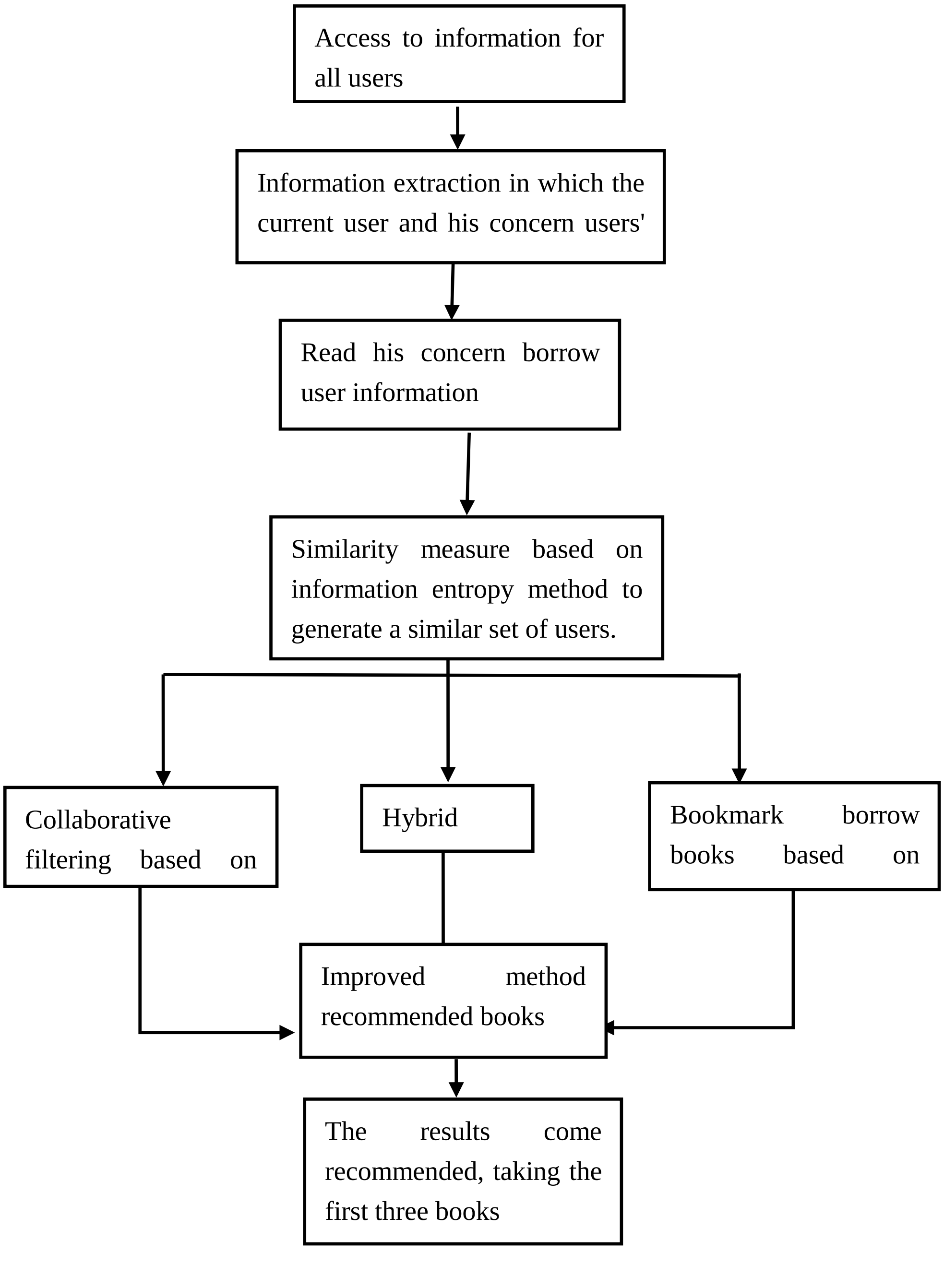

**5.3.5 Finally, some of the data from the title will be give into calculation, to verify**

**the correctness of the model**

This book is recommended quality evaluation system, can fully measure the level of quality through whether its predictions are accurate, because of different evaluation methods used in recommender systems are different. In order to verify the quality of the article constructed recommendation system, the paper title given data were brought into the test. Since this paper is collaborative filtering method, so you can give the book recommendation system repeatedly hitrate tests. (Source code, see Annex)

Data collection

Test data is the data used accessories given in the title, because the amount of data is very large attachments, can not all be brought into the calculations, so here is just one part of the data will be carried out into the calculations.

Hitrate calculated as:

$$hr(ui) = \frac{|T_{ui} \cap Xui|}{|T_{ui}|}$$

$T_{ui}$ is the result of which represents the results of the recommended books, $X_{ui}$ recommendation system use given herein recommended book data. Followed the adoption of the Law on hitrate K values constantly tested and found K value of 5 when the effect is better, therefore, will be assigned to 5 K, tested its Hitrate@n test results are as follows:

By comparing the test results, hitrate found results significantly better than the use of the sorted values unsorted Hitrate higher, Book recommendations and the results are related. Filtering effect based on books bookmarks and the results is nearly the same.[6]

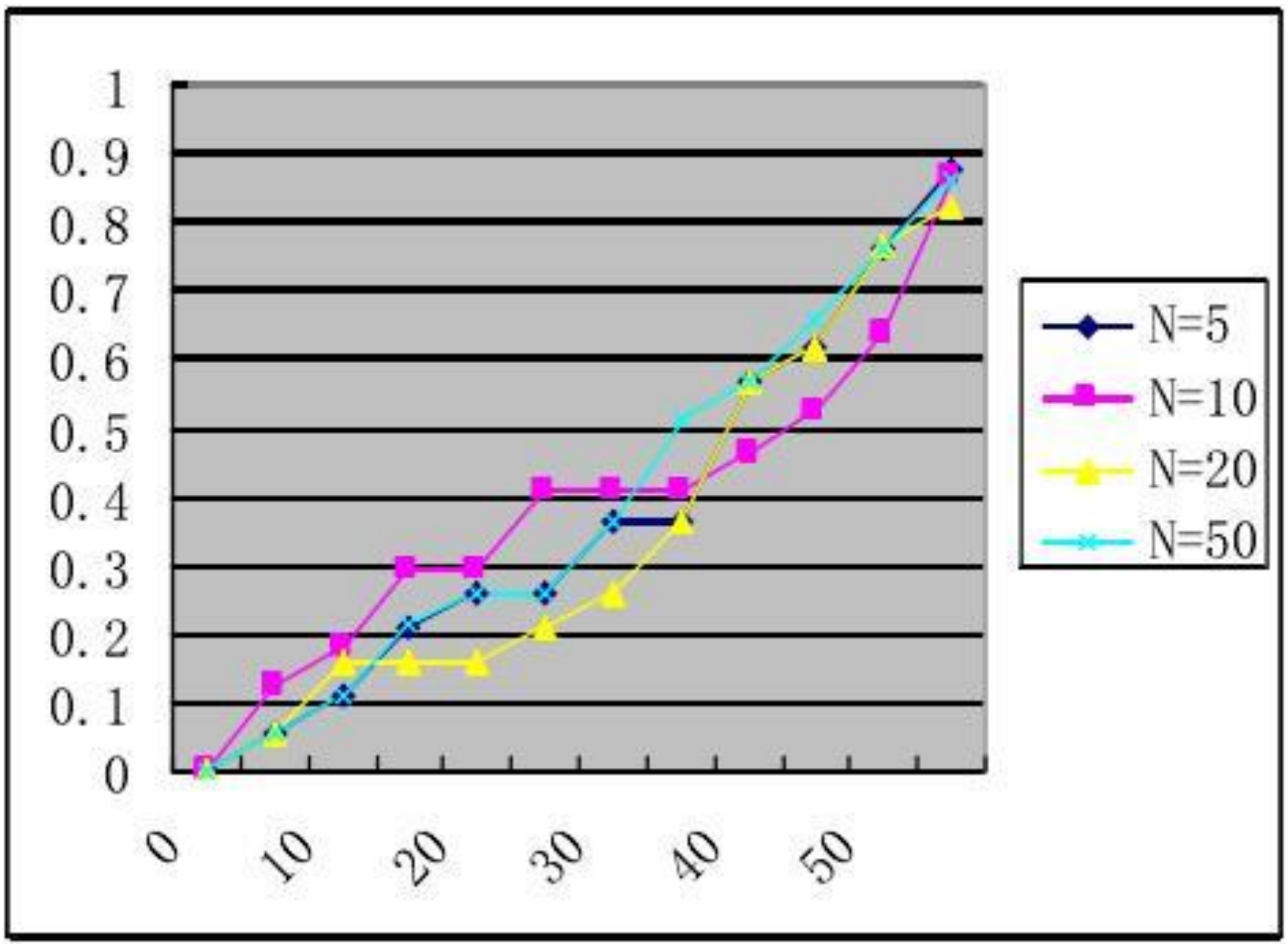

Inside, n represents the number of back to result, N indicates the continuous range of values of comprehensive sequencing.

The hybrid recommendation model has higher accuracy than the front, in the process of test, the effect of the mixed recommended and the value of N is not related, which proved that the paper book recommendations adopted by the model is effective.

## 6. Evaluation of the model

In this paper, the mathematical model adopted in the first question is AHP fuzzy comprehensive evaluation model. This method reduced the subjectivity caused by using AHP method, only makes the results more objective and persuasive.

Method in the second asked one used the data extraction and analysis, this method has some shortcomings, and was improved, the method is mainly used for the ant clustering forecasting model, this model has the advantages of cluster visualization in the calculation of similarity and by using the methods of the cosine Angle method, improves the accuracy of user ratings.

In question 3 method of using the model mainly in the collaborative filtering method based on information entropy again on the basis of the use of user similarity based on user ratings filtering and collaborative filtering based on books bookmark together build mixed collaborative filtering model for library evaluation, this method makes up the defect of the traditional methods, such as in high-dimensional sparse data, users pay close attention to the size of the circle intersection between much smaller and inconsistent, the traditional similarity measure methods cannot solve the similar situation, the results are more likely to decrease or increase the real similarities between users. So make the results more objective and accurate.

## 7. Improvement of the model

(1) Because of time and considering the data file is bigger, so there is no into the calculation carried out to verify all the data.

(2) On the recommendation of books, but also can comprehensively consider other factors, and other information users, and can improve the intelligent degree of recommended books system.